\documentstyle [12pt] {article}
\title {GENERALIZED LAPLACE METHOD FOR SIMPLE DETERMINATION
OF KERR-NEWMAN BLACK HOLE HORIZON RADIUS}

\author{Vladan Pankovi\'c$^{\ast,\sharp}$,
Simo Ciganovi\'c$^\sharp$, Jovan Ivanovi\'c$^\sharp$\\
$^\ast$Department of Physics, Faculty of Sciences, 21000 Novi
Sad,\\ Trg Dositeja Obradovi\'ca 4. , Serbia, vdpan@neobee.net \\
$^\sharp$Gimnazija, 22320 Indjija, Trg Slobode 2a, Serbia \\}

\date {}

\begin{document}
\maketitle

\vspace {0.5cm} PACS number: 04.70. -s \vspace {0.5cm}

\begin {abstract}
In this work we present a generalized Laplace method for a formal,
simple, quasi-classical, determination of the outer and inner
horizon radius of Kerr-Newman black hole. We consider classical
gravitational interaction between a thin, with homogeneously
distributed mass and electric charge, spherical (black) shell and
a probe particle. Also, we use relativistic equivalence principle.
Finally we suppose that probe particle propagates radially to
shell with speed of light while tangentially it rotates in common
with shell, so that total energy of a probe particle equals zero.
\end {abstract}

\vspace{0.5cm}

     As it is well-known [1] Laplace determined radius $R$ of a static black star, identical to horizon radius of Schwarzschild black hole, by simple, classical method. He supposed that static black star holds mass $M$ homogeneously distributed over its volume. Further, he supposed a probe particle with mass $m$ at black star surface, interacting classically gravitationaly with black star. Finally, he supposed that probe particle total energy, i.e. sum of gravitational potential energy and kinetic energy by probe particle radial (in respect to black star surface) propagation with speed of light $c$ equals zero, i.e.
\begin {equation}
         E_{tot}= \frac {mc^{2}}{2}- \frac {GmM}{R} = 0
\end {equation}
where $G$ represents Newtonian gravitational constant. In natural
system of units ($G=c=1$) expression (1) turns out in
\begin {equation}
         E_{tot}= \frac {m}{2}- \frac {mM}{R} = 0
\end {equation}
that implies
\begin {equation}
         R = 2M           .
\end {equation}
It is exactly equivalent to horizon radius of Schwarzschild black
hole with equivalent mass M [2].

In this work we shall present a generalized Laplace method for a
formal, simple, quasi-classical, determination of the outer and
inner horizon radius of Kerr-Newman black hole. We shall consider
classical gravitational interaction between a thin, with
homogeneously distributed mass and electric charge, spherical
black shell and a probe particle. Also, we shall use relativistic
equivalence principle. Finally we shall suppose that probe
particle propagates radially to black shell with speed of light
while tangentially it rotates in common with black shell, so that
total energy of a probe particle equals zero.

So, suppose that there is a thin, rotating around an axis that
holds center, spherical black shell, with radius $R$ and constant
angular velocity $\Omega$ as well as with homogeneously
distributed mass $M$ and electrical charge $Q$.

Suppose that at the spherical black shell there is a probe
particle with mass $m$ that rotates in common with black shell
(without relative rotation in respect to black shell) with
constant angular velocity $\Omega$. Also suppose that given probe
particle propagates radially (in respect to black shell surface)
with speed of light $c$.

Then, as it is well-known [3], classical kinetic energy of probe
particle at black shell surface is given by expression
\begin {equation}
         E_{k}= \frac {mc^{2}}{2} + \frac {m}{2} (\frac {j}{mr})^{2}
\end {equation}
where
\begin {equation}
    j=m\Omega^{2}R
\end {equation}
represents the probe particle angular momentum.

Further, classical potential energy of attractive gravitational
interaction between probe particle and black shell equals
\begin {equation}
          V_{g} = - \frac {GmM}{R}            .
\end {equation}

Since black shell is electrically charged with charge $Q$
homogeneously distributed over black shell surface, then, as it is
well-known [3], classical potential energy of the electrostatic
repulsive self-interaction of black shell is given by expression
\begin {equation}
          V_{c} =  \frac {1}{2}\frac {1}{4 \pi \epsilon_{0}} \frac {Q^{2}}{R}
\end {equation}
where $\epsilon_{0}$ represents the vacuum dielectric constant.
According to relativistic equivalence principle mass equivalent to
$V_{c}$ equals
\begin {equation}
          M_{c} = -\frac {V_{c}}{c^{2}} = - \frac {1}{2}\frac {1}{4\pi \epsilon_{0}}\frac {Q^{2}}{R}\frac {1}{c^{2}}            .
\end {equation}
It can be pointed out that mass $M_{c}$ is negative since
potential energy $V_{c}$ is positive, i.e. since electrostatic
self-interaction of black shell is repulsive.

Suppose now that there is classical gravitational interaction
between part of black shell with mass $ M_{c}$ and probe particle
so that potential energy of this interaction is
\begin {equation}
          V_{gc} = - \frac {GmM_{c}}{R} =  Gm\frac {1}{2}\frac {1}{4\pi \epsilon_{0}}\frac {Q^{2}}{{R}^{2}} \frac {1}{c^{2}}           .
\end {equation}

Then total potential energy of interaction between probe particle
and black shell equals
\begin {equation}
        V = V_{g} + V_{gc} =
        - Gm \frac {M}{R} + Gm \frac {1}{2} \frac {1}{4\pi \epsilon_{0}} \frac {Q^{2}}{{R}^{2}} \frac {1}{c^{2}}           .
\end {equation}

Finally, suppose that total energy of the probe particle equals
zero, i.e.
\begin {equation}
 E_{tot} = E_{k} + V_{g} + V_{gc} =
 \frac {mc^{2}}{2} + \frac {m}{2} (\frac {j}{m R})^{2}- Gm \frac {M}{R} + Gm \frac {1}{2} \frac {1}{4\pi \epsilon_{0}} \frac {Q^{2}}{R^{2}} \frac {1}{c^{2}} = 0  .
\end{equation}

In natural system of units ($G=c=4\pi \epsilon_{0}=1$) expression
(11) turns out in
\begin {equation}
 E_{tot}  = \frac {m}{2} + \frac {m}{2} (\frac {j}{mR})^{2}- m \frac {M}{R} + m\frac {1}{2}\frac {Q^{2}}{{R}^{2}} = 0  .
\end {equation}

As it is not hard to see the following is satisfied
\begin {equation}
   \frac {j}{mR}  = \frac {mvR}{mR} = \frac {m\Omega R^{2}}{mR} = \frac {M\Omega R^{2}}{MR}= \frac {J}{MR}
\end {equation}
where
\begin {equation}
     J= M\Omega R^{2}
\end {equation}
represents the angular momentum of the black shell.

Equation (12), according to (13), turns out in
\begin {equation}
           E_{tot}=  \frac {m}{2} + \frac {m}{2}(\frac {J}{MR})^{2} - \frac {mM}{R} + \frac {m}{2} \frac {Q^{2}}{R^{2}}= 0  .
\end {equation}
It, after simple transformations, implies
\begin {equation}
  \frac {R}{2} + \frac {1}{2}(\frac {J}{M})^{2} \frac{1}{R} - {M} + \frac {1}{2}\frac {Q^{2}}{R}= 0
\end {equation}
and
\begin {equation}
 M = \frac {R}{2}+ \frac {1}{2}(\frac {J}{M})^{2} \frac{1}{R}+ \frac {1}{2}\frac {Q^{2}}{R}= \frac {R}{2}+ \frac {1}{2}{{\it a}^{2}}{R}+ \frac {1}{2}\frac {Q^{2}}{R}
\end {equation}
\begin {equation}
 R_{\pm}= M \pm (M^{2} - (\frac {J}{M})^{2}- Q^{2})^{\frac {1}{2}}= M \pm (M^{2} - {\it a}^{2}- Q^{2})^{\frac {1}{2}}
\end {equation}
where
\begin {equation}
  {\it a} = \frac {J}{M}                        .
\end {equation}
while $R_{\pm}$ represent solutions of the quadratic algebraic
equation over $R$
\begin {equation}
  R^{2}  - 2MR + [(\frac {J}{M})^{2} + Q^{2}] = R^{2}  - 2MR + [{\it a}^{2} + Q^{2}] = 0
\end {equation}
corresponding to (16).

As it is not hard to see, $R_{+}$ and $R_{-}$ represent
expressions formally equivalent to expressions for radius of the
outer and inner horizon of Kerr-Newman black hole with equivalent
parameters $M$, $Q$ and $J$, i.e. ${\it a}$ [4].

In this way it is shown that radius of the outer and inner horizon
of Kerr-Newman black hole can be obtained formally by a simple,
quasi-classical method representing a generalization of Laplace
method for determination of the horizon radius of Schwarzschild
black hole. We considered classical gravitational interaction
between a thin, with homogeneously distributed mass and electric
charge, spherical (black) shell and a probe particle. Also, we
used relativistic equivalence principle. Finally we supposed that
probe particle propagates radially to shell with speed of light
while tangentially it rotates in common with shell, so that total
energy of a probe particle equals zero.

\section {References}

\begin {itemize}

\item [[1]] V. Stephani, {\it La Place, Weimar, Schiller and the Birth of Black Hole Theory}, gr-qc/0304087
\item [[2]] H. Goldstein, {\it Classical mechanics} (Addison-Wesley Co., Cambridge, 1953)
\item [[3]] R.P. Feynman, R. B. Leighton, M. Sands, {\it The Feynman Lectures on Physics}, Vol. 2, (Addison-Wesley Pub. Co. Inc., Reading, Mass., 1964)
\item [[4]] R. M. Wald, {\it General Relativity} (Chicago University Press, Chicago, 1984)

\end {itemize}

\end {document}